\documentclass[preprint,11pt,showpacs,final]{revtex4}

\usepackage{amsmath,amssymb}
\usepackage{amsfonts,amsthm}
\usepackage{graphics}
\usepackage{graphicx}
\usepackage{dcolumn}
\usepackage{color}
\usepackage{bm}
\begin{document}
\title{Frustration-induced quantum phase transitions in a quasi-one-dimensional ferrimagnet: Hard-core boson map and the Tonks-Girardeau limit}
\author{R.~R. Montenegro-Filho}
\email{rene@df.ufpe.br}
\author{M.~D. Coutinho-Filho}
\email{mdcf@ufpe.br}
\affiliation{Laborat\'orio de F\'{\i}sica Te\'orica e Computacional, Departamento de F\'{\i}sica, Universidade Federal de Pernambuco, 50670-901, Recife-PE, Brazil}
\begin{abstract}
We provide evidence of a superfluid-insulator transition (SIT) of magnons 
in a quasi-one-dimensional quantum ferrimagnet with {\it isotropic} competing
antiferromagnetic spin interactions. This SIT occurs between two distinct ferrimagnetic phases
due to the frustration-induced closing of the gap to a magnon excitation. It thus causes 
a coherent superposition of singlet and triplet states at lattice unit cells and 
a power-law decay on the staggered spin correlation function along the transverse direction to the 
spontaneous magnetization. A hard-core boson map suggests that asymptotically close to the
SIT the magnons attain the Tonks-Girardeau limit. The quantized nature of the 
condensed singlets is observed before a first-order transition to a singlet magnetic spiral phase 
accompanied by critical antiferromagnetic ordering. In the limit of strong frustration, 
the system undergoes a decoupling transition to an isolated gapped two-leg ladder and a 
critical single linear chain.    
\end{abstract}
\pacs{75.10.Pq,75.10.Jm,75.40.Mg,75.30.Kz,75.50.Gg}
\maketitle
\section{Introduction}
Recently, several experimental and theoretical studies indicate that, under very special conditions, 
magnons \cite{Magnons,Ruegg,MagnonQE} and polaritons \cite{Polaritons} undergo Bose-Einstein condensation (BEC) 
in two- and three-dimensional materials. In magnetic systems, BEC of magnons can be driven by an applied 
magnetic field ($h$) (Ref. \cite{Magnons}), by varying the external pressure \cite{Ruegg}, or by microwave pumping \cite{MagnonQE} 
In 1D gapped antiferromagnets, e. g., spin-1 chains \cite{Affleck} and single spin-1/2 two-leg ladders \cite{CoupledLadders}, the gap to the magnon excitation closes at a critical value ($h_c$) of the field and the magnetization increases as $(h-h_c)^{1/2}$. Although, {\it stricto sensu}, there is no BEC of magnons in these 1D systems, it is very appealing to describe the transition in terms of the condensation of the uniform component of the magnetization along the applied field \cite{Affleck}. In fact, rigorous results \cite{Pitaevskii} on low dimensional ($D\leq 2$) {\it uniform} interacting boson systems preclude the occurrence of BEC in finite temperature ($T$). In 2D systems phase fluctuations have mainly a thermal origin, so that only the $T=0$ condensate survives, with superfluid behavior persisting up to the Kosterlitz-Thouless temperature. 
In contrast, in 1D boson systems phase fluctuations have a quantum origin and 
there is no BEC, even at $T=0$, but superfluidity is expected \cite{Pitaevskii}. However, in {\it finite} 
systems the scenario is more complex, since in real confined systems \cite{Pitaevskii,Snoke2006} one may be
 dealing with metastable states.

In this work we introduce an {\it isotropic} Heisenberg spin Hamiltonian with two competing antiferromagnetic (AF)
exchange couplings [$J_1 (\equiv 1)$ and $J$] exhibiting a continuous quantum phase transition 
at a critical value $J_{c1}$ which, we argue, is a superfluid-insulator transition (SIT) of magnons
 associated with the creation of a coherent superposition of singlet and triplet states at lattice unit cells.
For $J=0$, the model shares its phenomenology and unit cell topology
with quasi-one-dimensional ferrimagnetic compounds \cite{Revisao}, such as
the line of trimer clusters present in copper phosphates \cite{trimers}, and 
the organic ferrimagnet PNNBNO (Ref. \cite{ORG}).
On the theoretical side, several features of the ferrimagnetic phase 
have been studied through Hubbard \cite{HUBBARD}, $t-J$ (Ref. \cite{Sierra}) 
 and Heisenberg \cite{HEISENBERG} models, including magnetic excitations \cite{OndadeSpin,PhysA} and the occurrence
  of new phases induced by hole doping of the electronic band \cite{Montenegro2006}.  
Also, the physical properties of the compound Cu$_3$(CO$_3$)$_2$(OH)$_2$ were successfully explained \cite{Kikuchi} 
by the distorted diamond chain model \cite{FrustTheor}, which is a system with three spin 1/2 magnetic sites
per unit cell and coupling parameters such that the ferrimagnetic state is frustrated. 

Numerical results have been obtained for finite clusters
through Density Matrix Renormalization Group (DMRG) (Refs. \cite{DMRG,error}) using
open boundary conditions and exact diagonalization (ED) using periodic boundary conditions 
and Lanczos algorithm. 

The paper is organized as follows: in Sec. \ref{Secao2} we introduce the model Hamiltonian 
and analyze the magnetic correlations of the competing phases close to $J=J_{c1}$. 
In Sec. \ref{Secao3} we define a hard-core boson model (HCB model), which is used to describe the 
main characteristics of the magnon SIT at $J=J_{c1}$, in particular, the Tonks-Girardeau limit. 
Further, in Sec. \ref{Secao4} we discuss the singlet magnetic spiral phase 
accompanied by critical antiferromagnetic ordering, which sets in after a first-order 
transition at $J=J_t$, as well as the decoupling transition, at $J=J_{c2}$, to an isolated gapped two-leg ladder and a 
critical single linear chain. Finally, a summary of the results is presented in Sec. \ref{Secao5}.
\section{Model Hamiltonian and Ordered Phases}
\label{Secao2}
\begin{figure}
\begin{center}
\includegraphics[width=0.41\textwidth,clip]{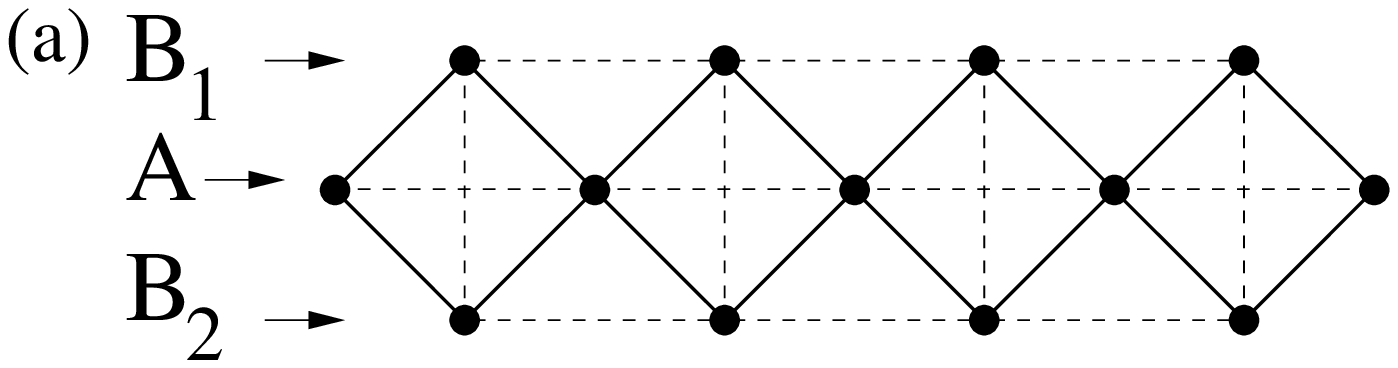}
\includegraphics[width=0.41\textwidth,clip]{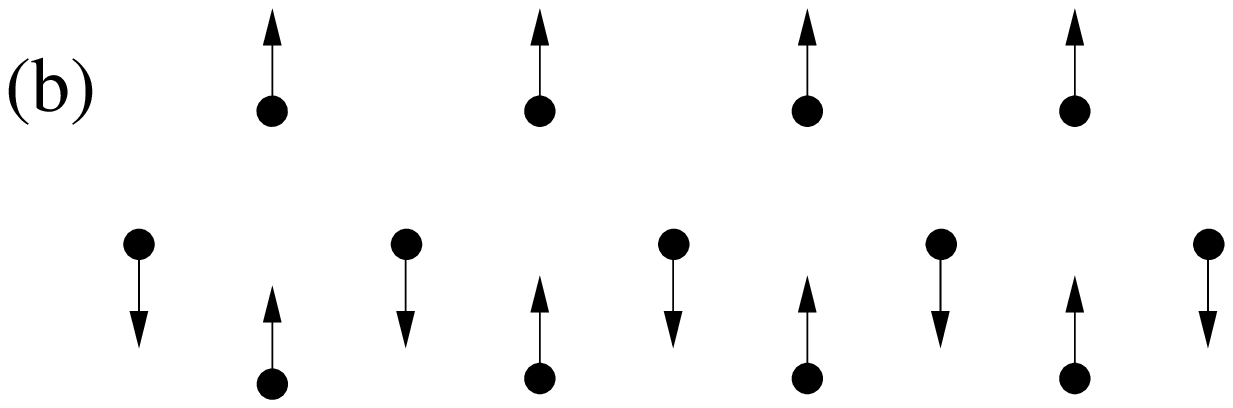}
\includegraphics[width=0.71\textwidth,clip]{Fig1c}
\caption{\label{Fig1} (a) Illustration of the A and B sublattices (circles)
and AF spin couplings
which favor (full lines) and destabilize (dashed lines) the LM
ferrimagnetic GS: $J_1 (\equiv 1)$ and $J$, respectively. (b) Illustration of the
LM ferrimagnetic GS. (c) Results (see text) for $S_g/S_{LM}$; dashed and dotted lines are guides to the eye.}
\end{center}
\end{figure}

The model Hamiltonian reads:
\begin{eqnarray}
&&H=\sum_{l=1}^{N_c}\sum_{\alpha=1,2}\mathbf{A}_l\cdot(\mathbf{B}_{\alpha l}+\mathbf{B}_{\alpha, l-1})+J(\sum_{l}\mathbf{A}_{l}\cdot\mathbf{A}_{l+1}\nonumber\\ 
&& +\mathbf{B}_{1l}\cdot\mathbf{B}_{2l}+\sum_{\alpha=1,2}\mathbf{B}_{\alpha,l}\cdot\mathbf{B}_{\alpha,l+1}),\label{ham}\\
\nonumber
\end{eqnarray}
as sketched in Fig. \ref{Fig1}(a). In Eq. (\ref{ham}), 
$\mathbf{A}_l$, $\mathbf{B}_{1l}$ and $\mathbf{B}_{2l}$ denote 
spin 1/2 operators at sites A$_{l}$, B$_{1l}$ and B$_{2l}$
of the unit cell $l$, respectively, and $N_c$ is the number of unit cells. 
For $J=0$ the model (named {\it AB$_2$ chain} or {\it diagonal ladder})
is bipartite and the Lieb-Mattis (LM) theorem \cite{LiebMattis} predicts a ground state (GS) total spin
 
\begin{equation}
S_g=\frac{|N_A-N_B|}{2}=\frac{N_c}{2}\equiv S_{LM}, 
\end{equation}
where $N_A$ ($N_B$) is the number of A (B$_1$ and B$_2$) sites. The GS 
spin pattern is represented in Fig. \ref{Fig1}(b).
In Fig. \ref{Fig1}(c) we report data for $S_g/S_{LM}$ 
as a function of $J$ using DMRG ($N_c=33$) and ED ($N_c=10$). Although the LM theorem is not applicable for $J\neq 0$, the ferrimagnetic phase ({\it F1 phase}) is robust up to $J\approx 0.342\equiv J_{c1}$ (Ref. \cite{Ivanov}), beyond which $S_g$ steadily decreases ({\it F2 phase}) before a first order transition to a phase with $S_g=0$ (apart from finite size effects) at $J\approx 0.445$. 
\begin{figure}
\begin{center}
\includegraphics[width=0.38\textwidth,clip]{Fig2a}
\includegraphics[width=0.38\textwidth,clip]{Fig2b}
\includegraphics[width=0.43\textwidth,clip]{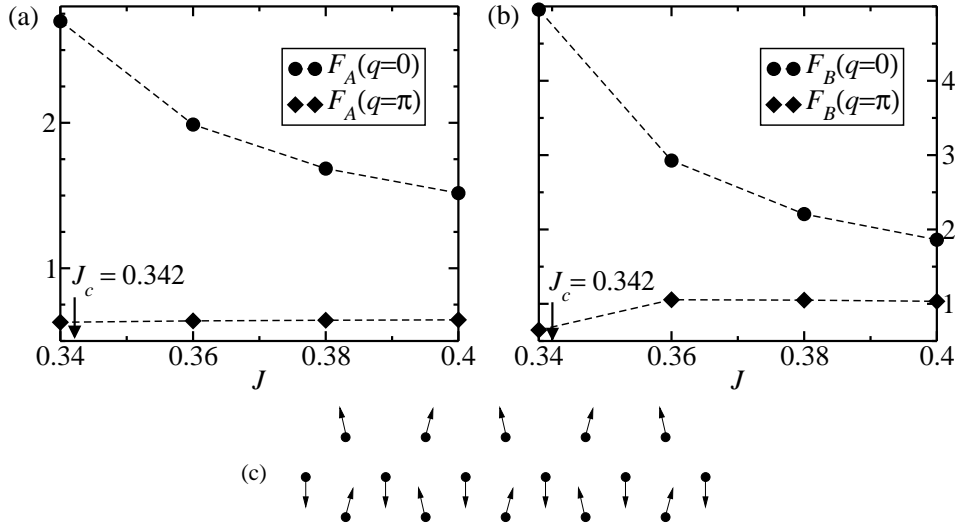}
\caption{\label{Fig2} DMRG results for the
magnetic structure factor, $F_X(q)$, at $q=0$ and $q=\pi$
for (a) $X=$ A and (b) $X=$ B (B$_1$ or B$_2$) spins
in a chain with $N_c=33$; dashed lines are guides to the eye.
(c) Illustration of the F2 phase.}
\end{center}
\end{figure}

In order to characterize the F2 phase, we have calculated the magnetic structure 
factor, 
\begin{equation}
F_X(q) =\sum_{l}^{N_c}C_{X}(l) \text{e}^{iql}, 
\end{equation}
with $q=2\pi n/(N_c-1)$, $n=0,1,...,N_c-1$, where $C_{X}(l)$ is the two-point correlation function between spins separated by $l$ unit cells at sites $X=A, B_1\text{ and } B_2$. 
We first noticed that the A spins remain ferromagnetically ordered as the critical point
$J_{c1}=0.342$ is crossed, although the magnitude of the peak at $q=0$ decreases for $J>J_{c1}$,
as displayed in Fig. \ref{Fig2}(a), while no peak is observed at $q=\pi$. 
The B$_i$ ($i=1\text{ or }2$) spins also remain ferromagnetically ordered (peak at $q=0$), with similar $J$-dependence, as shown in Fig. \ref{Fig2}(b). 
However, an extra peak at $q=\pi$ develops after the transition, which indicates the occurrence of a period-2 
modulation in the spin pattern for $J\gtrsim J_{c1}$.
Further, the average value of the 
correlation function $\langle \mathbf{B}_{1l}\cdot\mathbf{B}_{2l}\rangle$, 
which amounts to $\approx 0.25$ (triplet state) in the F1 phase, 
steadily decreases after the transition at $J_{c1}$. These findings suggest 
that the F2 phase would display a canted configuration, as illustrated in Fig. \ref{Fig2}(c).
However, to check whether these features are
robust in the thermodynamic limit, we have studied the finite size scaling 
behavior of the transverse (T) and longitudinal (L) order parameters in the F2 phase:
\begin{equation}
m_{X(L,T)}^2(q)=\frac{F_{X(L,T)}(q)}{N_c}, 
\end{equation}
\begin{figure}
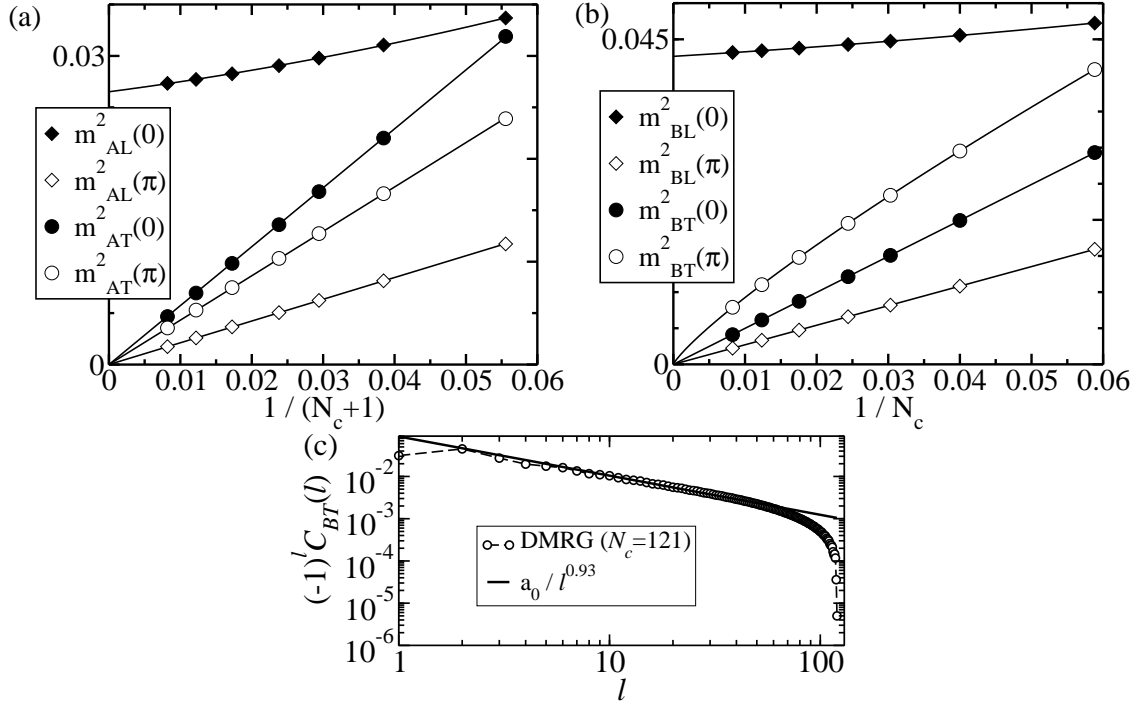

\begin{center}
\includegraphics[width=0.45\textwidth,clip]{Fig3a}
\includegraphics[width=0.45\textwidth,clip]{Fig3b}
\includegraphics[width=0.45\textwidth,clip]{Fig3c}
\caption{\label{Fig3} DMRG results for the square of the longitudinal (L) and transverse (T) order parameters at the spin se
ctor $S^z=S_g$ and $J=0.395$ for
(a) A  and (b) B (B$_1$ or B$_2$) spins (full lines are polynomial fittings).
 (c) DMRG results for the transverse staggered correlation function $C_{BT\pi}(l)$.}
\end{center}
\end{figure}
for $q=0$ (uniform component) and $q=\pi$ (staggered component), in the subspace of maximum total spin $z$-component ($S^z=S_g$).
The correlations are studied at $J=0.395$, for which $S_g=S_{LM}/2$, and
the results are shown in Fig. \ref{Fig3}. We confirmed that in the (extrapolated) thermodynamic limit the spins at sites A and B are ferromagnetically ordered, as indicated by $m^2_{XL}(q=0)\neq 0$ in Figs. \ref{Fig3}(a) and (b).
Further, since the A and B net magnetizations are oppositely 
oriented, the F2 phase is ferrimagnetic. 
The values of $m^2_{AL}(q=\pi)$, $m^2_{BL}(q=\pi)$ and $m^2_{BT}(q=0)$
nullifies linearly with system size, which evidences short-range correlations.
On the other hand, the best fitting to 
the data for $m^2_{BT}(q=\pi)$ presents a nonlinear dependence with the inverse of the system 
size and also nullifies in the thermodynamic limit. This behavior indicates that 
the staggered correlation function of the spins at sites B along the transverse direction
to the spontaneous magnetization,
$C_{BT\pi}(l)$, exhibits a power-law decay, as explicitly confirmed in Fig. \ref{Fig3}(c). 
We thus conclude that for $J_{c1}<J<0.445$ the GS is also ferrimagnetic 
but with critical correlations along the transverse direction to the spontaneous magnetization ({\it F2 phase}). 
\begin{figure}
\begin{center}
\includegraphics[width=0.50\textwidth,clip]{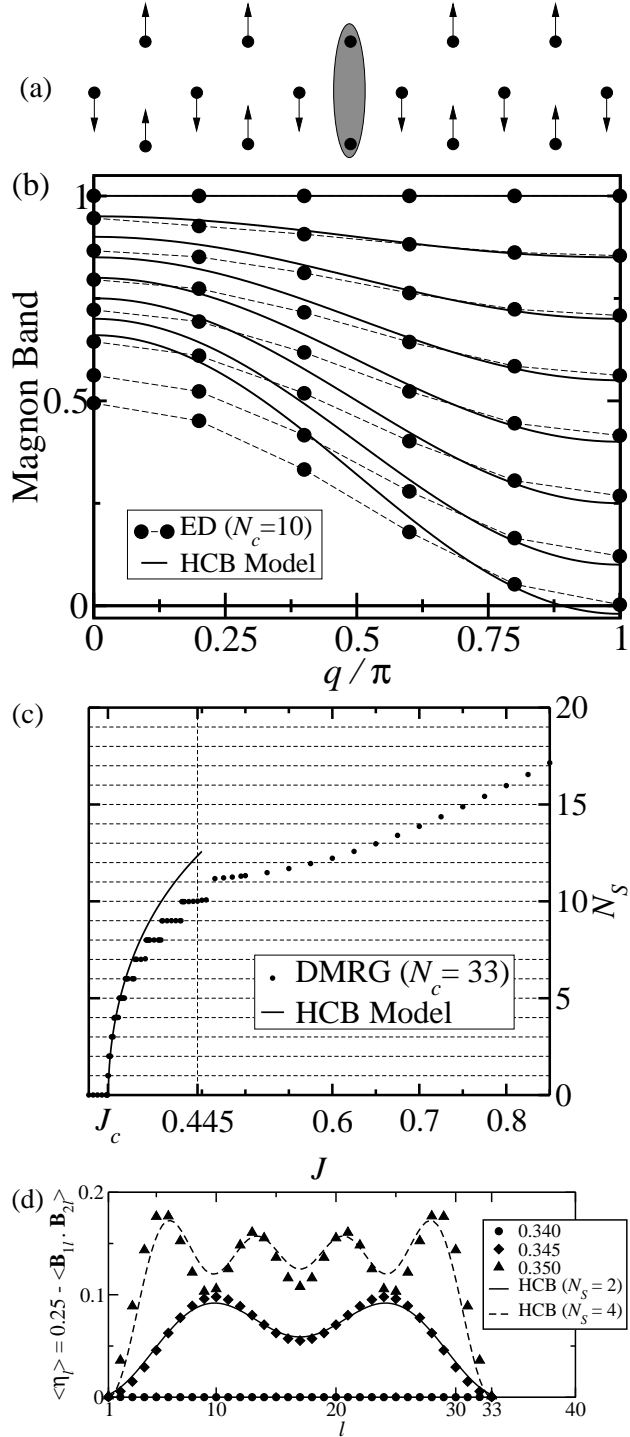}
\includegraphics[width=0.50\textwidth,clip]{Fig4b}
\includegraphics[width=0.50\textwidth,clip]{Fig4c}
\includegraphics[width=0.5\textwidth,clip]{Fig4d}
\caption{\label{Fig4}(a) Illustration of the relevant magnon excitation for $J=0$:
ellipse indicates a local singlet state. (b) Magnon band
for $J=0.00,0.05,0.10,0.15,0.20,0.25,0.30$ and $0.34$, from top
to bottom. (c) $N_S$ as a function of $J$. Full lines are the HCB model
predictions in the TG limit and dashed lines are guides to the eye.}
\end{center}
\end{figure}

Next we focus on the effect of $J$ on the magnetic excitations.
For $J=0$ the Hamiltonian exhibits three magnon modes \cite{PhysA,OndadeSpin}.
One is AF, i. e., the spin is raised by one unit with respect
to the GS total spin, while the other two are ferromagnetic, associated with the 
lowering of the GS total spin by one unit. The AF gapped dispersive mode is 
responsible for a quantized plateau in the magnetization curve as function of $h$
and should also exhibit condensation, as suggested by the numerical data 
in Ref. \cite{PhysA}.
One of the ferromagnetic magnons is the gapless dispersive Goldstone mode,
 while the other is a flat mode and is the relevant excitation for
the transition at $J=J_{c1}$. To understand some nontrivial features 
of this excitation, we must comment on the symmetry properties of the model.
For $J=0$ the Hamiltonian is invariant under the exchange of the B sites at the {\it same} cell. This symmetry implies that spins at these B sites can be found only in singlet or triplet states (mutually exclusive possibilities); in the GS only triplets are found. 
The relevant magnon is a localized gapped mode which induces the formation of a singlet pair in one cell, as illustrated in Fig. \ref{Fig4}(a). For $J\neq0$, 
this local symmetry is explicitly broken and the spins at these B sites 
can be found in a coherent superposition of singlet and triplet states.
In Fig. \ref{Fig4}(b), data using ED for the magnon band ($q=2\pi n/N_c, n=0,1,...,N_c-1$) is displayed for various values of $J$ before the transition point. For $J=0$ the band is flat with a gap $\Delta_0\approx 1.0004$. 
By increasing $J$, the bandwidth increases and the gap to the GS lowers, closing
at the wave vector $q=\pi$ at the transition point.

\section{Hard-Core Boson Model, Superfluid-Insulator Transition and the Tonks-Girardeau Limit}
\label{Secao3}
The GS total number of singlets is given by 
\begin{equation}
N_S=\sum_{l=1}^{N_c} \langle \eta_l\rangle,
\end{equation} 
with singlet density $\langle \eta_l\rangle= \langle s^\dagger_l s_l\rangle$, where  
\begin{equation}
s^\dagger _l\equiv\frac{1}{\sqrt{2}}(B_{1l,\uparrow}^\dagger B_{2l,\downarrow}^\dagger-B_{1l,\downarrow}^\dagger B_{2l,\uparrow}^\dagger)
\end{equation}
is the creation operator of a singlet pair at cell $l$ and 
$B_{i l,\sigma}^\dagger$ is the creation operator of an electron with spin $\sigma$ 
at the B$_i$ ($i=1, 2$) site of cell $l$. In fact, it is easy to show that 
\begin{equation}
\langle\eta_l\rangle=\frac{1}{4}-\langle\mathbf{B}_{1l}\cdot\mathbf{B}_{2l}\rangle,
\end{equation}
so $N_S=0$ for $J=0$.
In Fig. \ref{Fig4}(c) we observe that $N_S$ starts to increase in steps of unity 
after $J=J_{c1}$, indicating the quantized nature of the condensing singlets.

We now examine the nature of the quantum critical point at $J=J_{c1}$. 
For this purpose we split the Hamiltonian of Eq. (\ref{ham}) in three terms:
the first favors ferrimagnetism,
\begin{equation}
H_{AB}=\sum_{l}\mathbf{A}_l\cdot(\mathbf{S}_{l}+\mathbf{S}_{l-1}),
\label{hamab}
\end{equation}
where $\mathbf{S}_l=\mathbf{B}_{1l}+\mathbf{B}_{2l}$; the second one 
favors AF ordering between A spins, i. e.,
\begin{equation}
H_{A}=J\sum_l\mathbf{A}_l\cdot\mathbf{A}_{l+1},
\end{equation}
and shall play no significant role in our analysis; 
the last term, also unfavorable to ferrimagnetism, 
is a two-leg ladder Hamiltonian connecting spins at sites $B_1$ and $B_2$ (discarding a constant factor) \cite{Fouet2006,Hikihara2001}:
\begin{equation}
H_{B}=\frac{J}{2}\left (\sum_l S_l^2+\sum_l \mathbf{S}_l\cdot \mathbf{S}_{l+1}
+\sum_l \mathbf{D}_l\cdot \mathbf{D}_{l+1} \right) ,
\end{equation}
where $\mathbf{D}_l=\mathbf{B}_{1l}-\mathbf{B}_{2l}$.  
We represent the Hamiltonian 
in a basis with two states for each pair $\mathbf{B}_{1l}$ and $\mathbf{B}_{2l}$: 
the singlet and the triplet component in the magnetization direction. 
 In addition, we define the vacuum of the HCB model as the state with this triplet component in each cell. We now study 
the GS energy when a number $N_S$ of singlet pairs is added to the vacuum.
For $J=0$, the energy cost of a singlet pair is $\Delta_0$ (the 
gap to the flat mode); thus, for $N_S$ singlets, the contribution
from $H_{AB}$ is $N_S\Delta_0$. The first term in $H_{B}$ is 
diagonal and will add a factor of $-JN_S$; the second causes a 
repulsion between singlets and adds also an extra factor of $-JN_S$; finally, the last term in $H_{B}$ 
introduces the singlet itinerancy. Grouping these contributions, we arise 
to a model of hard-core bosons with nearest-neighbor repulsion:

\begin{equation}
H_{S}=(\Delta_0-2J)N_S+\frac{J}{2}\sum_{l}\eta_l\eta_{l+1}+
\frac{J}{2}\sum_l(s^\dagger_ls_{l+1}+h.c.). 
\end{equation}

We remark that the hard-core boson interaction is implied by the algebra of the singlet operators: 
\begin{eqnarray}
{[}s_l,s^{\dagger}_l{]}_{+}&=&1; \mbox{ and} \\
{[}s_l,s_m{]}_{-}&=&0 \mbox{ for } l\neq m. 
\end{eqnarray}
Before the transition, the single magnon dispersion relation,
\begin{equation}
\omega_q(J)=\Delta_0-2J+J\cos{q}, 
\end{equation}
agrees well with the numerical data for $q\approx \pi$, as can be seen in Fig. \ref{Fig4}(b).
The resulting critical point: $\omega_{q=\pi}(J_{c1,S})=0$, i. e., 
\begin{equation}
J_{c1,S}=\frac{\Delta_0}{3}\approx 0.333, 
\end{equation}
is in excellent agreement with the 
numerical prediction $J_{c1}=0.342$. Moreover, the closing of the magnon gap is also
in excellent agreement with the prediction 
\begin{equation}
\Delta_J=\omega(q=\pi)=3(J_{c1,S}-J)
\end{equation} 
and with the expected linear vanishing of the Mott gap \cite{Fisher}: $z\nu=1$, where $z=2$ and $\nu=1/2$ are the correlation length and dynamic critical exponents, respectively (see below). 

After the transition and in the highly diluted limit ($\frac{N_S}{N_c}\equiv \eta\rightarrow 0$), the energy of $N_S$ hard-core bosons in 1D is well
approximated by the energy of $N_S$ free spinless fermions \cite{Affleck}. Through
this map, the energy density reads:
\begin{eqnarray}
\mathcal{E}_{GS}(J)&=&\frac{E_{GS}(J)}{N_c}=\int_{-k_F}^{k_F}\frac{dk}{2\pi}[\epsilon_{k}(J)-\mu_F]\\
                   &\approx& 3(J_{c1,S}-J)\eta+\frac{J\pi^2\eta^3}{6}, 
\end{eqnarray}
where $k_F=\pi\eta$ and 
\begin{equation}
\epsilon_k(J)-\mu_F=\omega_{k+\pi}(J)\approx -3(J-J_{c1,S})+\frac{J k^2}{2}. 
\end{equation}

\begin{figure}
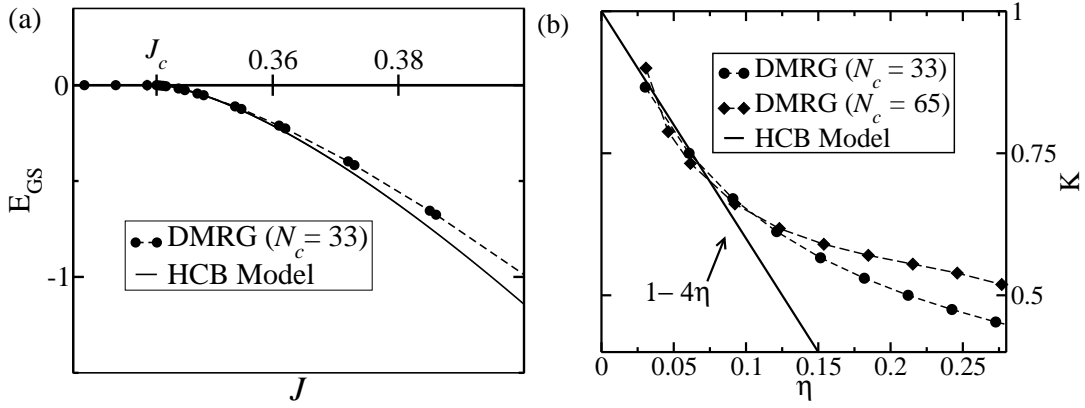

\begin{center}
\includegraphics[width=0.42\textwidth,clip]{Fig5a}
\includegraphics[width=0.44\textwidth,clip]{Fig5b}
\caption{\label{Fig5}(a) Ground state energy, $E_{GS}$, relative to the energy of the LM state.
(b) Luttinger Liquid exponent, $K$, as function of $\eta$.
Full lines are the HCB model predictions in the TG limit and
dashed lines are guides to the eye.}
\end{center}
\end{figure}

Notice that the Fermi chemical potential satisfies the Tonks-Girardeau (TG) limit 
\cite{TonksGirardeau,Pitaevskii,TonksGas} 
(1D Bose gas of impenetrable particles), corresponding to an infinitely high repulsive 
potential in the Lieb-Liniger solution \cite{LiebLiniger} of the $\delta$-function 1D Bose gas:
\begin{equation}
\mu_F=\epsilon_F(J)=\frac{\pi^2 J\eta^2}{2}, 
\end{equation}
where $J^{-1}$ is the fermion mass, $\hbar\equiv 1$, 
and $\eta$ is the density of singlets for $J\geq J_{c1,S}$ derived from the equilibrium
condition $\partial_{\eta}\mathcal{E}_{GS}(J)=0$:
\begin{equation}
\eta=\frac{\sqrt{6(J-J_{c1,S})}}{\pi\sqrt{J}}\text{, $\eta\rightarrow 0$,}
\label{eta}
\end{equation}
much in analogy with the 1D field-induced transition. Further, in Fig. \ref{Fig4}(d) we display the good agreement between the numerical estimate for the density 
$\langle \eta_l\rangle$ of a two (four) particle state, $N_S=2$ ($N_S=4$), in an open system and the HCB model in a continuum space given by \cite{Fouet2006}
\begin{equation}
\langle\eta(l)\rangle=\frac{2}{N_c-1}\sum_{n=1}^{N_S} \sin^2(k_n l), 
\end{equation}
with $k_n=1,...,\frac{\pi N_S}{N_c-1}$. Also, as shown in Figs. \ref{Fig1}(c), \ref{Fig4}(c) and \ref{Fig5}(a), 
the HCB model predictions for 
\begin{equation}
\frac{S_g}{S_{LM}}=1-2\eta, 
\end{equation}
$\eta$ and $E_{GS}(J)$, respectively, are very close to the numerical data for $J\gtrsim J_{c1}\approx J_{c1,S}$.

On the other hand, using the Luttinger liquid description \cite{Voit,affleckSTAT} for our highly diluted HCB model 
we have the following general relations for the sound velocity $c$ and the compressibility $\kappa$:
\begin{eqnarray}
c & = & \frac{\pi J \eta}{K};  \\
\frac{1}{\eta^2\kappa}& = & \frac{\pi c}{K},
\end{eqnarray}
where $K$ is the Luttinger parameter 
governing the decay of the correlation functions. However, since
\begin{equation}
\frac{1}{\eta^2\kappa}=\frac{d^2\mathcal{E}_{GS}}{d \eta^2}=\pi^2 J\eta,
\end{equation} 
it implies that $K=1$; thus $c=\pi J \eta=Jk_F$, in accord with the 
TG limit \cite{Pitaevskii,TonksGas}. Further, taking $\eta$ as the order parameter of the SIT, Eq. (\ref{eta}) implies $\beta=1/2$, while $\eta^2 \kappa$ diverges with a critical exponent $\alpha=\gamma=1/2$, in agreement with the 
scaling and hyperscaling relations \cite{Fisher}: 
\begin{eqnarray}
\alpha+2\beta+\gamma&=&2,\\
2-\alpha&=&\nu(d+z), 
\end{eqnarray}
respectively, assuring that the SIT is in the free spinless gas universality class \cite{sachdev}.
 
In an interacting Bose gas \cite{gia2}, $K=1/2$  is the separatrix between
systems dominated by superfluid fluctuations, $K>1/2$, from those dominated by charge density
fluctuations, $K<1/2$,  (in our magnetic model spin fluctuations prevail). Affleck and collaborators \cite{AffleckScattering} have succeeded in taking into account corrections from interactions between pairs of dilute magnons parametrized by a scattering length, $a$, thus implying that 
\begin{equation}
K=1-2am+O(m^2),
\end{equation}
where $m$ is the 
field-induced magnetization for the $S=1$ chain, with $a\approx - 2$. The predicted increase 
of $K$ with $m$ was confirmed by numerical calculations \cite{AffleckScattering}. This parametrization can also be implemented in our problem. In fact, in Fig. \ref{Fig5}(b) we show 
that $K=1-4\eta$, with $a\approx 2$, fits quite well the data for the Luttinger liquid 
parameter in the highly diluted regime. $K$ was calculated using DMRG 
and assuming 
\begin{equation}
C_{BT\pi}\sim \frac{a_0}{l^{\frac{1}{2K}}}.
\end{equation}
\section{Spiral Correlations, Weakly Coupled AF Chains, and Ladder-Chain Decoupling}
\label{Secao4}

We now turn our attention to the transition point $J_{t}\approx 0.445$, which marks the onset of a singlet
phase, as can be seen in Fig. 1(c), characterized by non-quantized values of $N_S$, as shown in Fig. 4(c). 
On the other hand, from the Hamiltonians in Eqs. (8)-(10), we can infer that for $J>>1$ 
the system should decompose into a linear chain (A sites) and an 
isotropic two-leg ladder system (B$_1$ and B$_2$ sites); see Fig. 1(b). 
The linear chain is known 
to be gapless with critical spin correlations (power-law decay), while the two-leg ladder is gapped 
with exponentially decaying correlations. In what follows we discuss the complex phase diagram in the region $J>J_{t}$. 
\begin{figure}
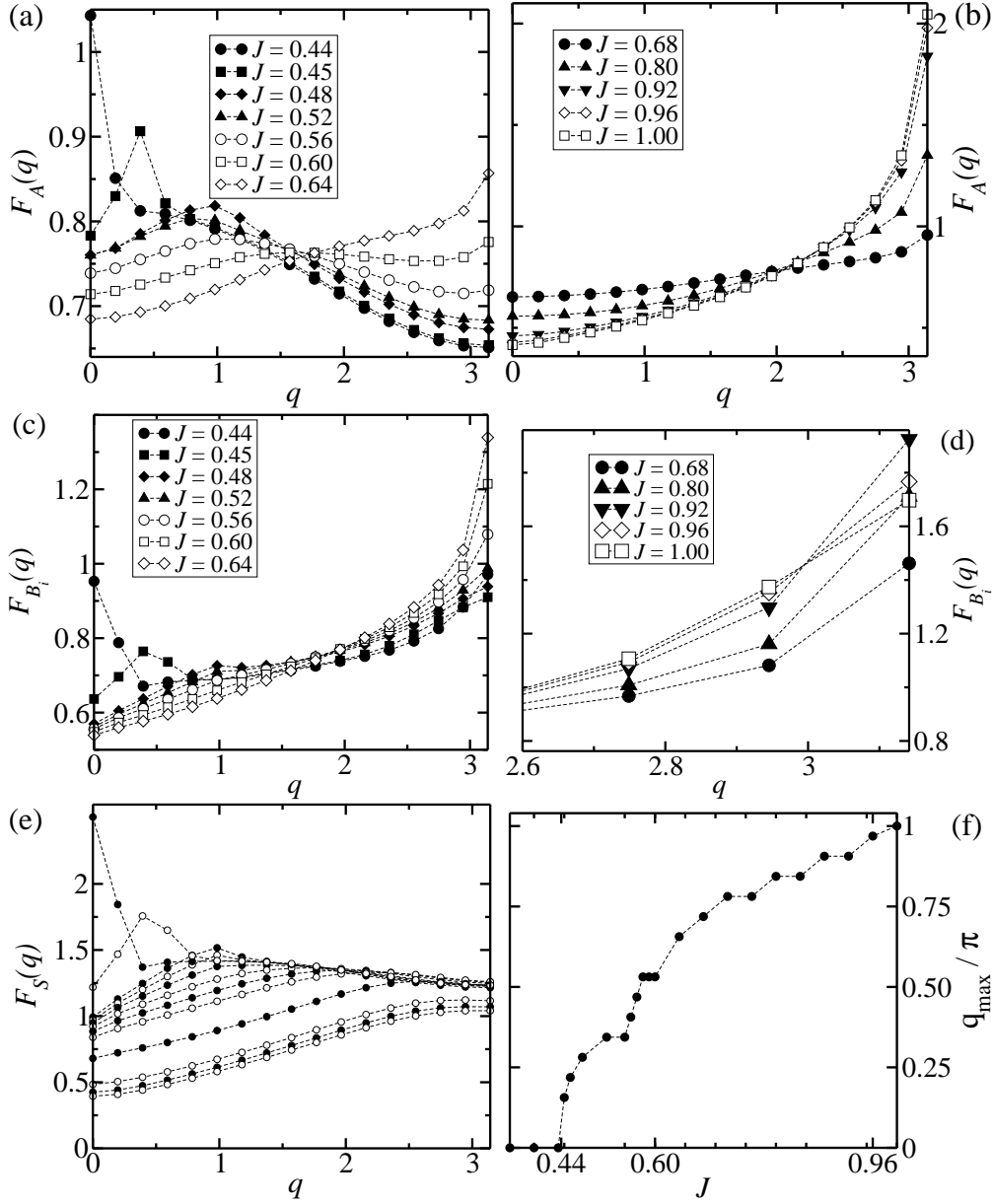

\begin{center}
\includegraphics[width=0.4\textwidth,clip]{Fig6a}
\includegraphics[width=0.4\textwidth,clip]{Fig6b}
\includegraphics[width=0.4\textwidth,clip]{Fig6c}
\includegraphics[width=0.4\textwidth,clip]{Fig6d}
\includegraphics[width=0.4\textwidth,clip]{Fig6e}
\includegraphics[width=0.4\textwidth,clip]{Fig6f}
\caption{\label{Fig6} Magnetic structure factor $F_X(q)$ for the
$A$ spins [(a) and (b)], $N_c=32$, and for the $B_i$ spins [(c) and (d) with $i=1$ or 2], $N_c=33$, for the
indicated values of $J$. (e) Magnetic structure factor for the composition
$\mathbf{S}_l=\mathbf{B}_{1l}+\mathbf{B}_{2l}$
and $J=0.44,0.45,0.48,0.52,0.56,0.60,0.64,0.68,0.80,0.92,0.96\text{~~and~~}1.00$,
from top to bottom at $q=0$. (f) The value of
the wave-vector for which the peak at the magnetic structure factor exhibited
in (e) is observed. Dashed lines are guides to the eye.}
\end{center}
\end{figure}

Initially, we display in Fig. 6 the magnetic structure factors $F_A(q)$ and $F_{B_i}(q)$, with $i=1\text{ or }2$, 
as well as $F_S(q)$, which is associated to the magnetic structure of the composite spin 
$\mathbf{S}_l=\mathbf{B}_{1l}+\mathbf{B}_{2l}$.
In Fig. 6(a) we see that $F_A(q)$ peaks at $q=0$ for $J=0.44$, i. e., the system remains in the F2 phase and the A spins are
ferromagnetically ordered. For $J=0.45$, a sharp peak in a spiral wave-vector $q_{max}$ is observed. The peak broadens 
and $q_{max}$ increases with increasing $J$. For $J=0.56$ we notice the emergence of a commensurate AF peak, 
coexisting with the spiral one, particularly for $J\geq0.60$, as seen in Figs. 6(a) and 6(b). 
On the other hand, we observe in Fig. 6(c) the presence of two peaks in $F_{B_i}(q)$ for $J=0.44$: the $q=0$ peak 
associated with the ferromagnetic ordering of the $B_i$ sites in the F2 phase, and the $q=\pi$ peak related to the 
critical staggered transverse correlation at the same phase. Likewise, for $J=0.45$, a spiral peak is observed at the same 
wave-vector $q_{max}$ of $F_A(q)$. Further, notice in Fig. \ref{Fig6}(d) that the magnitude of the AF peak drops in the 
interval $0.96\leq J<1.00$. 

In order to develop a physical meaning of the above referred data, we first point out that the coupling between spins 
at A and B sites occurs through the composition $\mathbf{S}_l=\mathbf{B}_{1l}+\mathbf{B}_{2l}$, as can be seen in 
Eq. (8). Further, as the singlet component of $\mathbf{S}_l$ is magnetically inert, only its triplet 
components affect the magnetic ordering at the A sites. In fact, as shown in Figs. 6(e) and 6(f), 
short-range spiral ordering is observed in the magnetic structure of $\mathbf{S}_l$ up to $J\approx 1.00$. 
However, since the peak is weak and broad for $J\gtrsim0.6$, its feature is 
overcomed by the AF one in the data of Figs. 6 [(a)-(d)]. In the sequence, we focus on the AF 
ordering observed for $J\gtrsim 0.6$ and study how the system approaches the ladder-chain decoupling.
\begin{figure}
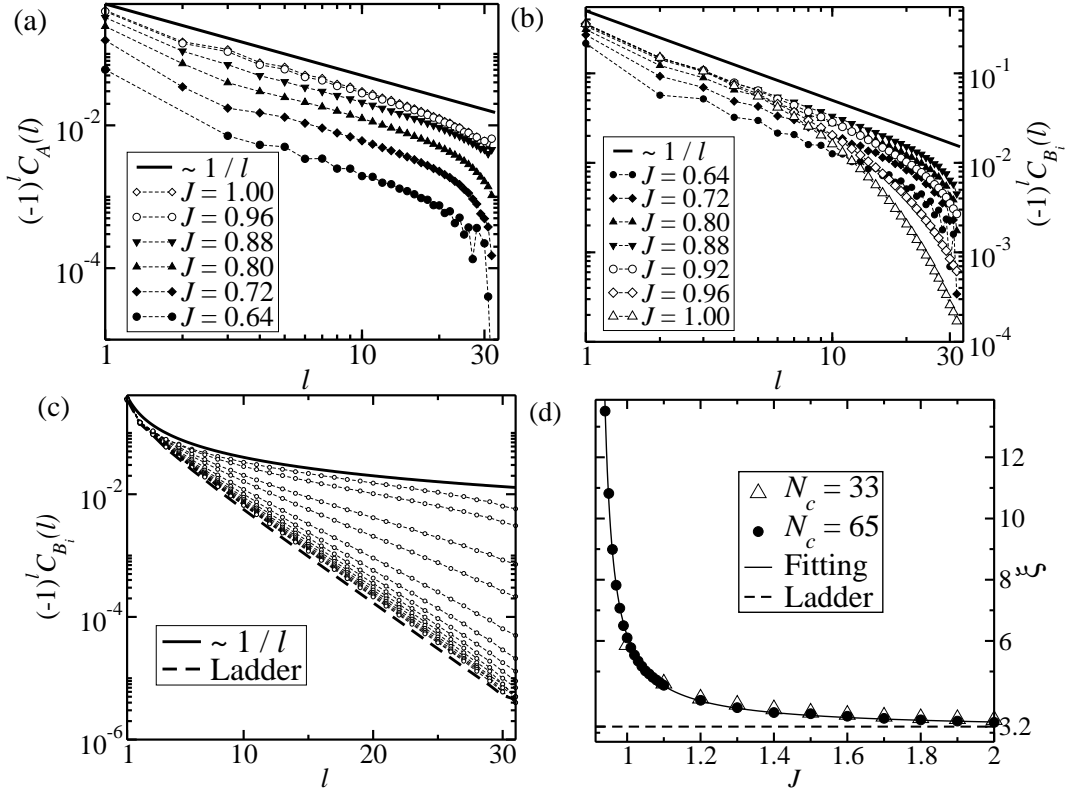

\includegraphics[width=0.4\textwidth,clip]{Fig7a}
\includegraphics[width=0.45\textwidth,clip]{Fig7b}
\includegraphics[width=0.4\textwidth,clip]{Fig7c}
\includegraphics[width=0.42\textwidth,clip]{Fig7d}
\caption{\label{Fig7}Staggered correlation functions between (a) A spins, $N_c=32$, and (b) $B_i$ (with $i=1$ or 2) spins,
$N_c=33$, for the indicated values of $J$. In (a) and (b) solid lines indicate the asymptotic behavior for
a single chain. (c) Staggered correlation functions
between $B_i$ (with $i=1$ or 2) spins for $J=0.88,0.92,0.96,1.0,1.1,1.2,1.3,1.4,1.5,1.6,1.7,1.8,1.9\mbox{ and }2.0$,
circles-dash from top to bottom and $N_c=33$. For comparison we plot the behavior for a single chain (solid line)
and for a two-leg ladder with 32 rungs.
(d) Correlation length $\xi$ as a function of $J$: solid line indicates
the fitting of the data for $N_c=65$ to Eq. (\ref{nlsm}) with $g$
given by Eq. (\ref{gfinal}); dashed line indicates the value of $\xi$ ($\approx 3.2$) for a two-leg ladder.}
\end{figure}

In Fig. 7(a), we present the staggered AF correlation function between A-spins as $J\rightarrow 1$. 
As observed, its behavior is well described by that found in a single linear chain, 
which is asymptotically given by 
\begin{equation}
C(l)\sim \frac{(-1)^l}{l},
\label{linearbe}
\end{equation} 
apart from logarithmic corrections \cite{Voit}. A similar behavior is observed in Fig. 7(b) 
for the staggered AF correlation between $B_i$-spins up to $J=0.88$, a value beyond which the shape of the curve 
is visibly changed. In order to understand this dramatic behavior, we recall that in a two-leg ladder system 
the asymptotic form of the 
correlation is given by \cite{LadderCorr} 
\begin{equation}
C(l)\sim \frac{(-1)^l e^{-l/\xi}}{l^{1/2}},
\label{ladderbe}
\end{equation}
where $\xi(\approx 3.2, \text{see Ref.\cite{WhitePRL1994}})$ defines the correlation length, associated with the gapped 
spin liquid state of this system. Indeed, as displayed in Fig. 7(c) the staggered correlations $C_{B_i}$ 
asymptotically approaches the correlation in a two-leg ladder system. In Fig. 7(d) we present the behavior of $\xi$ as a function
of $J$ for $N_c=33$ and $N_c=65$. These data were obtained 
by a proper fitting of $C_{B_i}$ in the 
interval $l_0<l<(N_c/2)$:  
starting from $J=2$ and taking $l_0 \approx 6$ 
(about twice the value of $\xi$ of a two-leg ladder),  
we find $\xi$; twice this value of $\xi$ was used as input 
($l_0 = 2\xi$) for the 
next chosen value of $J$, and so on.
Moreover, we have obtained a good fitting to these data by using 
the two-loop analytic form of the O(3) 
non-linear sigma model (NLSM) correlation 
length in (1+1) dimension \cite{Brezin}:
\begin{equation}
\xi=a e^{\frac{2\pi}{g}}\left (1+\frac{2\pi}{g}\right )^{-1},
\label{nlsm}
\end{equation}
where $a$ is a constant and $g$ is the NLSM coupling. 
Further, we assume (see below) that  
the coupling $g$ is the one suitable to the 
anisotropic quantum Heisenberg two-leg ladder 
to the NLSM \cite{nlsm}:
\begin{equation}
g=2\kappa\sqrt{1+\frac{J_{\perp}}{2J_{\parallel}}},
\label{coupling}
\end{equation}
where $J_{\perp}$ ($J_{\parallel}$) is the exchange 
coupling between spins at the same rung (leg) and 
$\kappa$ is a constant that depends on the choice of the lattice regularization. 

In order to justify Eq.~(\ref{coupling}) for $g$, 
we consider a mapping of the model 
Hamiltonian, Eq. (\ref{ham}), to the Hamiltonian 
of an isolated two-leg ladder by eliminating the 
spin degrees of freedom associated with the A sites. 
The mapping is performed, in a semiclassical  
manner, by the following assumption on $H_{AB}$ [Eq. (\ref{hamab})]:
\begin{equation}
H_{AB}\rightarrow \overline{H}_{AB}=\gamma\sum_l{\mathbf{A}_l\cdot\mathbf{S}_l},
\end{equation}
where $\gamma$ is an effective coupling constant. 
This amounts to reduce the A-B coupling 
to spins within the same unit cell, and 
cell-cell interactions are taken into account 
through the effective coupling $\gamma$.
We now write: $(\mathbf{A}_l+\mathbf{S}_l)^2 = 
\mathbf{A}_l^2 + \mathbf{S}_l^2 + 2 \mathbf{A}_l \cdot \mathbf{S}_l$,
with $\mathbf{S}_l^2 = \mathbf{B}_{1l}^2 + \mathbf{B}_{2l}^2 
+ 2 \mathbf{B}_{1l} \cdot \mathbf{B}_{2l}$;  
since within a unit cell $(\mathbf{A}_l+\mathbf{S}_l)^2 \approx (1/2)^2$ 
in an AF phase, and dropping constant terms, 
$\overline{H}_{AB}$ can be written as
\begin{equation}
\overline{H}_{AB}=-\gamma\sum_l\mathbf{B}_{1l}\cdot\mathbf{B}_{2l}.
\end{equation}
Since correlations between spins at $A$ sites does not play a 
significant role close to the transition, we discard the term 
$H_A$, and, finally, obtain the following {\it anisotropic} two-leg ladder 
Hamiltonian:
\begin{equation}
H\rightarrow H_{aL}=\overline{H}_{AB}+H_B,
\end{equation}
where the exchange couplings are given by 
\begin{equation}
J_{\perp}=J-\gamma,
\label{jperp} 
\end{equation}
\begin{equation}
J_{\parallel}=J.
\label{jpar}
\end{equation}
Substituting Eqs. (\ref{jperp}) and (\ref{jpar}) into 
Eq. (\ref{coupling}), we find the effective NLSM coupling:
\begin{equation}
g=\kappa\sqrt{\frac{6(J-J_{c2})}{J}},
\label{gfinal}
\end{equation}
where $J_{c2}=\gamma/3$. 
We have fitted the data in Fig. 7(d) 
to Eq. (\ref{nlsm}), with $g$ given by
Eq. (\ref{gfinal}), and $a$, $\kappa$ and $J_{c2}$ as fitting parameters. 
The obtained value of $a$ (=2.7) 
is such that $\xi\rightarrow 3.1$ as $J\rightarrow\infty$, 
which agrees with the expected value 
for an isolated {\it isotropic} two-leg 
ladder ($\approx 3.2$), while $\kappa=4.5$ and $J_{c2}=0.91$, 
in agreement with the correlation function behavior shown in Fig. 7(b). 
\begin{figure}
\begin{center}
\vspace*{0.3cm}
\includegraphics[width=0.45\textwidth,clip]{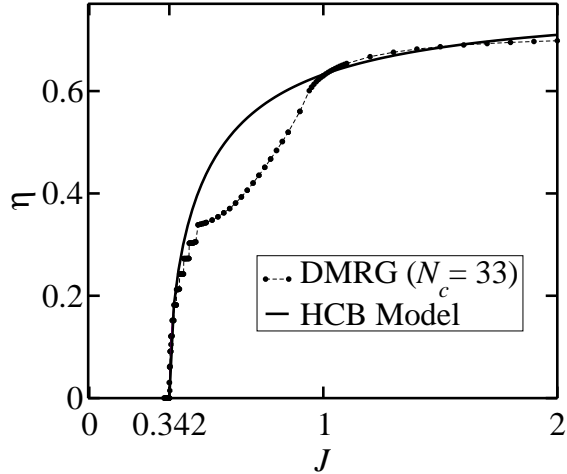}
\caption{\label{Fig8} Density of singlets as function of $J$.}
\end{center}
\end{figure}

Finally, in Fig. \ref{Fig8} we display the very interesting behavior 
of the density of singlets, $\eta$, as function of $J$. It is clear that
the effect of the A-spins and singlet-singlet interaction is relevant only 
for $J_t \lesssim J \lesssim J_{c2}$, otherwise the  solution, Eq. (21), for 
low density of singlets can be extended to the region of low density 
of triplets above $J_{c2}$ (strongly coupling limit), where correlations between B-spins 
are exponentially small [see Eq. (32)]. 
In fact, the asymptotic value predicted by Eq. (21), i. e., 
$\eta=\sqrt{6}/\pi\approx 0.78$, compares well with the numerical one: $\approx 0.71$. 
\section{Summary and Conclusions}
\label{Secao5}
\begin{figure}[h]
\begin{center}
\includegraphics[width=0.70\textwidth,clip]{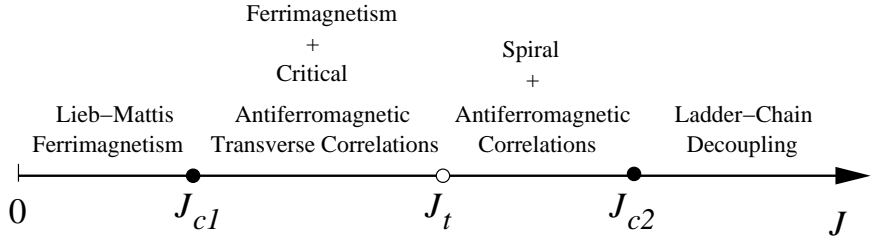}
\caption{\label{Fig9} Schematic representation of the phase diagram.}
\end{center}
\end{figure}
In this work we have derived 
the rich phase diagram of a three-leg spin Hamiltonian related to 
quasi-one-dimensional ferrimagnets, as function of a frustration parameter $J$  
which destabilizes the ferrimagnetic phase. 
In Fig. \ref{Fig9} we present 
an illustration of the obtained phase diagram, 
which displays two critical points, 
$J_{c1}\approx 0.342$ and $J_{c2}\approx 0.91$, 
and a first order transition point 
at $J_t\approx 0.445$.
Through DMRG, exact diagonalization and a hard-core boson model, 
we have characterized the transition at $J_{c1}$ 
as an insulator-superfluid transition 
of magnons (built from the coherent superposition of singlet and triplet states 
between B sites at lattice unit cells), with a well defined Tonks-Girardeau limit 
in the high diluted regime.
Ferrimagnetism with critical staggered correlations 
in a direction transverse to the spontaneous magnetization 
is observed for $J_{c1}<J<J_{c2}$.
Further, for $J_{c1}<J<J_{t}$ the number of singlets 
in the lattice is quantized, 
while above the first order transition at $J=J_{t}$
this quantity is a continuous one.
Also, in the interval $J_{t}<J<J_{c2}$ the magnetic 
structure factor displays a singlet phase 
with incommensurate ($q\neq 0\text{ and }\pi$) spiral 
and AF peaks.
However, the spiral peak broads and the AF peak is the salient 
feature as $J$ increases within this phase. 
At $J=J_{c2}$ a remarkable gapped two-leg ladder / critical single-linear chain 
decoupling transition occurs, 
characterized by an essential singularity in the correlation 
length as predicted by the NLSM through  
a mapping of our model onto an anisotropic quantum Heisenberg two-leg ladder.  
For $J \gg J_{c2}$ the ladder approaches the isotropic limit 
(full decoupling), while the linear chain remains critical. 

In summary, our reported results clearly reveal that 
frustrated quasi-one-dimensional 
magnets are quite remarkable systems to study magnon condensation,
including the crossover to coupled ladder systems of 
higher dimensionality \cite{orignac}
and related challenging phenomena \cite{challenging},
as well as frustration-driven quantum decoupling transition 
in ladder systems.

\section{Acknowledgments}
\label{Secao6}
We acknowledge useful discussions with A. S. F. Ten\'orio and E. P. Raposo. 
This work was supported by CNPq, Finep, FACEPE and CAPES (Brazilian agencies). 

\bibliography{frustbib}

\begin{thebibliography}{40}
\bibitem[{err()}]{error}
\bibinfo{note}{In the DMRG calculation we have retained from 300 to 1080 states
  per block. The discarded density matrix weight ranges from 10$^{-10}$ to
  10$^{-7}$, typically 10$^{-8}$.}
\end{thebibliography}
\end{document}